%% file: conference_101719.tex
\documentclass[conference]{IEEEtran}
\IEEEoverridecommandlockouts
\usepackage{cite}
\usepackage{amsmath,amssymb,amsfonts}
\usepackage{algorithmic}
\usepackage{graphicx}
\usepackage{textcomp}
\usepackage{xcolor}
\usepackage{hyperref}
\usepackage{float} 
\usepackage{array}
\usepackage[linesnumbered,ruled,vlined]{algorithm2e}
\usepackage{enumitem}

\def\BibTeX{{\rm B\kern-.05em{\sc i\kern-.025em b}\kern-.08em
    T\kern-.1667em\lower.7ex\hbox{E}\kern-.125emX}}
\begin{document}

\title{Traffic Intersection Simulation Using Turning Movement Count Data in SUMO: A Case Study of Toronto Intersections\\

\thanks{This work was supported by the Natural Sciences and Engineering Research Council (NSERC) of Canada.}
}

\author{\IEEEauthorblockN{Harshit Maheshwari}
\IEEEauthorblockA{\textit{Faculty of Business \& IT} \\
\textit{Ontario Tech University}\\
Oshawa, Canada \\
harshit.maheshwari@ontariotechu.ca}
\and
\IEEEauthorblockN{Li Yang}
\IEEEauthorblockA{\textit{Faculty of Business \& IT} \\
\textit{Ontario Tech University}\\
Oshawa, Canada \\
li.yang@ontariotechu.ca}
\and
\IEEEauthorblockN{Richard W. Pazzi}
\IEEEauthorblockA{\textit{Faculty of Business \& IT} \\
\textit{Ontario Tech University}\\
Oshawa, Canada \\
richard.pazzi@ontariotechu.ca}

}

\maketitle 

\begin{abstract}
Urban traffic simulation is vital in planning, modeling, and analyzing road networks. However, the realism of a simulation depends extensively on the quality of input data. This paper presents an intersection traffic simulation tool that leverages real-world vehicle turning movement count (TMC) data from the City of Toronto to model traffic in an urban environment at an individual or multiple intersections using Simulation of Urban MObility (SUMO). The simulation performed in this research focuses specifically on intersection-level traffic generation without creating full vehicle routes through the network. This also helps keep the network's complexity to a minimum. The simulated traffic is evaluated against actual data to show that the simulation closely reproduces real intersection flows. This validates that the real data can drive practical simulations, and these scenarios can replace synthetic or random generated data, which is prominently used in developing new traffic-related methodologies. This is the first tool to integrate TMC data from Toronto into SUMO via an easy-to-use Graphical User Interface. This work contributes to the research and traffic planning community on data-driven traffic simulation. It provides transportation engineers with a framework to evaluate intersection design and traffic signal optimization strategies using readily available aggregate traffic data.
\end{abstract}

\begin{IEEEkeywords}
Intersection modeling, SUMO, Urban Traffic Simulation \end{IEEEkeywords}

\section{Introduction}\label{Introduction}
Intersections are nodes where two or more traffic streams converge and conflict, making them an essential component in the road infrastructure network. They often become bottlenecks \cite{yuan2014identification}, leading to congestion and safety concerns due to their limited capabilities in terms of movements (left, right, and through) and in managing various transportation mediums, including but not limited to pedestrians, cars, buses, trucks, and bikes. Accurate modeling of intersections can offer valuable insights into improving and devising new algorithms for various traffic control methods. One of the most frequently researched methods is traffic signals optimization \cite{eom2020traffic}, where Machine Learning (ML) based methods are proven to be performing better than regular \cite{kumar2024adaptive} due to how they adapt to data in real-time, but they require realistic modeling of the traffic scenario, which depends on using the right tools along with quality real-world data which depict variable driving conditions \cite{zhao2024survey}. For instance, while optimizing traffic signals using Reinforcement Learning (RL), the data quality used to train the agent is directly proportional to how well it adapts when deployed in the real world \cite{kamal2024digital}. Traditionally, however, a significant challenge in simulation studies has been obtaining quality input data that depict real-world scenarios. Many researchers resort to synthetic data, generally random trips due to lack of better data, to devise and test their algorithm. This may result in unrealistic scenarios \cite{shokrolah2022turning}\cite{essa2016comparison}\cite{zambrano2016using}, resulting in good results while testing but subpar results during implementation. In traffic engineering, the “garbage-in, garbage-out” principle applies\cite{lieberman2014brief}: random or oversimplified input will result in random or unreliable output. This is especially true for critical input data like traffic volumes, movement data, and origin-destination patterns. 

In this work, we present a novel tool, CrossFlow, that uses real-world turning movement counts (TMC) data from the City of Toronto \footnote{ \href{https://open.toronto.ca/}{https://open.toronto.ca}} to streamline the process of generating intersection traffic scenarios in SUMO. The proposed tool automates the conversion of raw empirical data into simulation-ready inputs, simplifying what is typically a tedious manual process. The simulation reflects authentic demand patterns by using actual vehicle turn volume (vehicle turning right, left, and going through) aggregated into 15-minute bins. This study aims to demonstrate that synthetic data can be replaced with real-world data and depict close to actual traffic conditions. Using real turn count data also allows direct verification: if the simulation is properly set up, the number of vehicles passing through the intersection in the simulation should correspond closely to the observed counts within the expected variability. 

The data-driven approach presented in this paper via this tool improves the fidelity of results and makes the simulation setup more accessible and efficient for practitioners as they can more quickly generate scenarios reflective of realistic scenarios without advanced programming knowledge. Such ease of use and the accuracy of using real-world data in virtual scenarios enhance the approach's applicability in real-world planning. To the best of our knowledge, this is the first work that proposes a tool to integrate real TMC data from Toronto into SUMO via an automated and user-friendly interface. The contributions of this paper are threefold:
\begin{enumerate}
\item{We propose a methodology for converting intersection movement data into a format usable by a microscopic simulator, ensuring that all approach and turn movements are represented accurately.}
\item{A Python-based tool is proposed with an easy-to-use graphical user interface (GUI) that streamlines the process of scenario creation, network configuration, and running SUMO simulations for individual or multiple intersections. The source code for CrossFlow is made publicly available on GitHub\footnote{ \href{https://github.com/ANTS-OntarioTechU/CrossFlow}{https://github.com/ANTS-OntarioTechU/CrossFlow}} for any researcher and practitioner to use and develop.}
\item{We evaluate CrossFlow using real data from Toronto, comparing simulated traffic volumes with the original turn counts to assess accuracy. We also present potential use cases to illustrate how traffic engineers and researchers can apply this tool for intersection analysis and planning.}
\end{enumerate}

The remainder of this paper is organized as follows: Section \ref{Background} briefs about urban traffic simulation tools, discusses the need for using real data in simulations, and reviews related works. Section \ref{Methodology} outlines the process followed to convert TMC data into SUMO-compatible data. Section \ref{Tool} details the tool's architecture and features, including the GUI and how it interfaces with SUMO. Section \ref{Evaluation} presents various traffic simulation scenarios and discusses the tool's capabilities in generating those scenarios. Section \ref{Use} describes multiple use cases and applications that can be explored with this tool. Section \ref{Conclusion} wraps up the paper with key findings and future directions in Urban Traffic Simulation.

\section{Background and Related Works}\label{Background}
Synthetic data can limit how it depicts driver behavior; it cannot capture the intricacies and randomness of human drivers. Using real-world data to enhance the simulation quality has gained traction in recent years. Studies have shown that using default or uncalibrated parameters for microsimulation can yield poor results compared to observed data \cite{essa2016comparison}. Without proper adjustment, the initial simulation overestimates traffic volume, as presented in \cite{zambrano2016using} by over 200\%, which was further fixed by employing an iterative heuristic method to tune the input until the simulated volumes aligned closely with real values. This underlines the limitation of synthetic and unguided approaches and highlights the need for robust integration techniques to import real data into microsimulation models. Instead of relying on randomly generated or synthetic data, researchers have explored methods to directly integrate observed traffic data by extracting vehicle volumes and trajectories and converting them into microsimulation-compatible models.

Modern urban traffic simulation platforms provide a robust environment to model and analyze vehicle interactions. Commercial software like Paramics\cite{Cameron1996}, VISSIM\cite{Fellendorf2010}, AIMSUN \cite{Casas2010}, and open-source tools like SUMO\cite{Lopez2018}, CARLA\cite{Dosovitskiy2017}, MATSim\cite{Horni2016}, and CityFlow\cite{Zhang2019} are widely used to simulate traffic in an urban environment. Each simulator comes with its own set of advantages, depending on the task \cite{saidallah2016comparative}. Most of these simulators allow vehicle trajectory simulation at an individual scale for an urban intersection. These simulators benefit from using real data, including vehicle approach volumes or turning counts, to produce meaningful results. SUMO has emerged as one of the leading open-source traffic simulation platforms capable of modeling inter-modal traffic systems, including road vehicles, public transport, and pedestrians. It features an extensive library of prepackaged and external tools for simulation tasks, and its continuous open-source development makes it well-suited to our current task. Integrating actual TMC data into microsimulation improves realism and enables the model to serve as a virtual testbed for intersection improvements. By mirroring actual demand patterns, the simulation can identify the most critical approaches or turning movements and evaluate targeted interventions like signal timing changes or added turn lanes under true-to-life conditions.

TMC data extraction and integration techniques have gained traction in recent studies to improve the realism of urban traffic simulation. TMC data consists of vehicle count and movement direction, which can be classified into left, right, and through the intersections. Typical TMC data is recorded as an aggregate of traffic volumes every 15 minutes. TMC data is a key aspect of intersection design and planning, informing decisions on signage placement and traffic signal installation \cite{10185495}. In practice, this type of data is widely used by transportation planners to determine lane configurations, signal phasing, and timing at junctions. Therefore, a realistic simulation of an intersection hinges on accurate TMC inputs that reflect true demand on each movement. Obtaining high-quality TMC data can be challenging, depending on the method deployed. Traditionally, it involves manual counting \cite{karunathilake2024cn+}, which can be labor-intensive. Many new research methods employ computer vision (CV) techniques to analyze vehicle trajectories and calculate the volume using the intersection's permanent video feeds or drones \cite{bock2020ind}. The authors of \cite{shirazi2023intersection} obtain traffic data for their study by developing a CV-based system to estimate vehicle movement data from a video feed and then import those counts to SUMO. By calibrating the simulation using real intersection designs and counts, they were able to evaluate traffic signal performance with high fidelity to actual conditions. Their results highlighted that even detailed metrics, including lane occupancy and travel time, closely matched when demand input came from observed data. Another method to record vehicle movement data is to employ vehicle detector sensors at the incoming and outgoing edges of the intersection. Still, not all sensors support tracking the vehicle's unique signature, resulting in accurate incoming and outgoing vehicle counts with inaccurate or non-Movement data within an intersection.

Several studies have focused on implementing TMC data in simulators to analyze intersection performance and optimize Traffic Signal Control (TSC). The authors in \cite{shokrolah2022turning} identified the critical edge in the intersection network by analyzing the movement causing a high delay and then allocating more green time to these critical turning movements, showing a measurable improvement in vehicle travel times after TSC optimization. Beyond offline analysis, there is a growing trend in real-time simulations that use live traffic count data. \cite{wang2023real} proposes a study where TMC data is calculated in real-time using a video feed from Unmanned Aerial Vehicles (UAV). The goal of such a system is to create an up-to-date simulation that mirrors real-world traffic scenarios, and the authors were able to get a little over 90\% overall accuracy, theoretically enabling them to advance applications like dynamic signal timing adjustments and incident management in a virtual environment. These studies have demonstrated that incorporating TMC data can significantly improve the realism of urban intersection modeling.

In summary, the literature strongly indicates that using real traffic volume data can enhance the simulation and bridge the gap between simulation and reality. Many existing studies have documented the pitfalls in the traditional methods of using synthetic or randomly generated data, which fails to match the intricacies of real-world traffic behaviors. These approaches require extensive calibration and tuning \cite{Lopez2018} \cite{flotterod2009cadyts} to match real-world data, which can be time-consuming and may still not accurately represent real-world conditions. In contrast, our proposed methodology uses direct implementation techniques to infuse real traffic data into simulation models. This approach mitigates the shortcomings of synthetic data. In addition to using real data, we streamlined the workflow by importing the data, generating a simulator-compatible map, and running the simulation without manual processing. The user can input unique Toronto intersection IDs, and CrossFlow can do the rest. 

\section{Methodology}\label{Methodology}
\subsection{Data Source: City of Toronto- Traffic Volumes}
The foundation of our simulation scenarios is the turn movement count data\cite{TorontoTrafficVolumes} obtained from the City of Toronto, which is available to the public via their Open Data Portal. The city continuously collects multi-modal intersection counts over 14-hour periods (for data after September 2023) at a single location. This data includes the volume of motor vehicles broken down by type (car, bus, and truck) along with their movement through the intersection (left-turn, right-turn, and through-movement for each leg of the intersection). Each count record is aggregated into 15-minute bins, which are accompanied by an intersection name and ID, time, and the coordinates for the intersection. The data also consists of bike and pedestrian volumes, but is currently out of scope for this study.

\subsection{Building the Intersection Network in SUMO}
    \label{buildMap}
\begin{figure}[tb]
    \centering
    \includegraphics[width=\columnwidth]{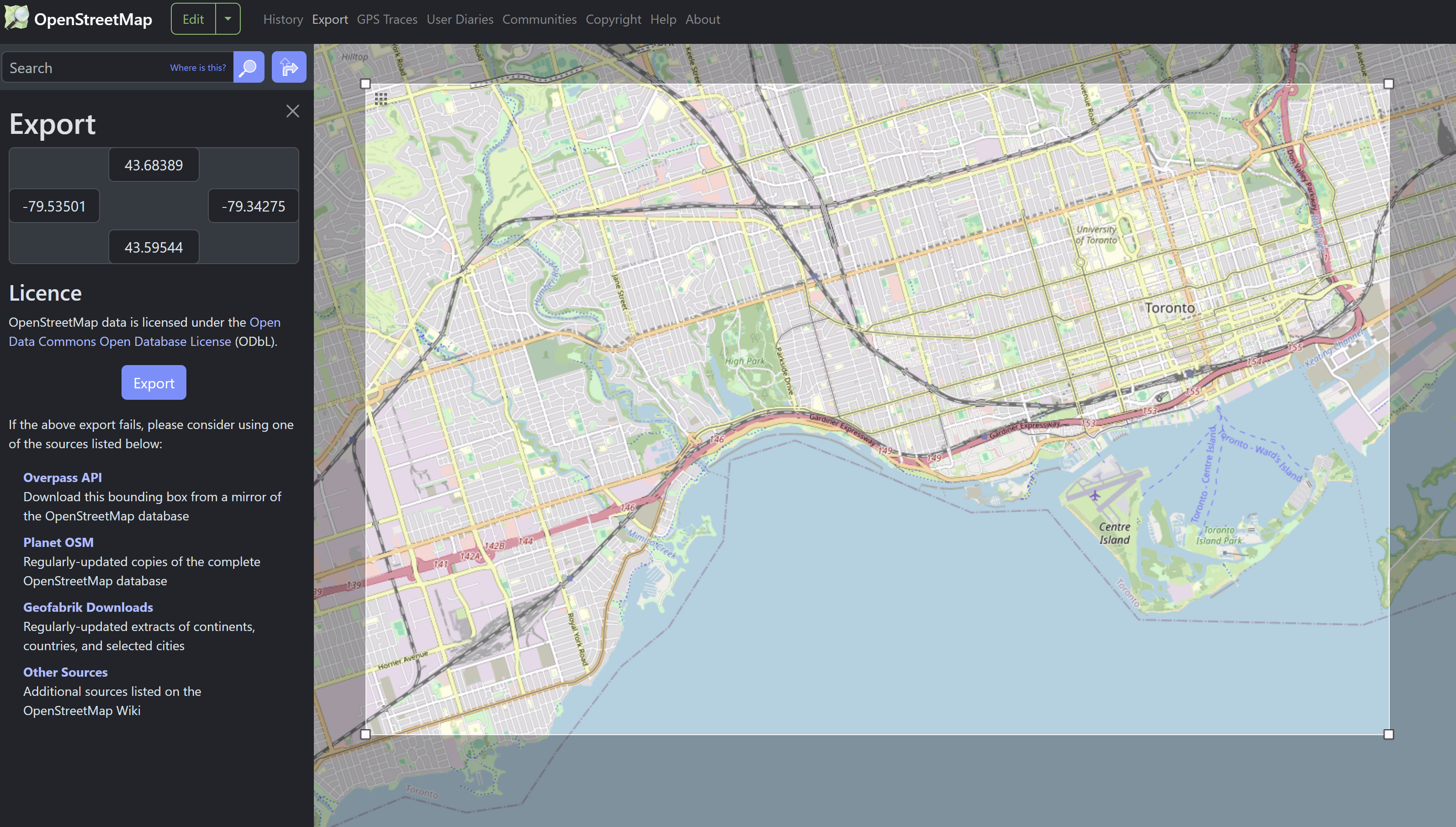}
    \caption{Open Street Maps Export Demonstration.}
    \label{fig:osm_export}
\end{figure}

Before injecting the data, the first step is to prepare the map file, which will simulate the network in SUMO. The OSM Web Wizard, which comes packaged with SUMO, allows the extraction of a small part of the map powered by OpenStreetMap for simulation purposes. Alternatively, OpenStreetMap offers an export option on their website, as shown in Figure \ref{fig:osm_export}, that can be saved locally. Once the map area is selected and exported, netedit can be used to modify the network topology if needed and export the map to SUMO-compatible format using netedit or netconvert, '.net.xml'. Both netedit and netconvert come prepackaged with SUMO as well. By constructing the network in SUMO-compatible format, we have a digital twin of the intersection where we can inject vehicles according to the observed counts.
Additionally, the tool features an 'Auto Fetch' option, which will automatically fetch the complete map area of the selected intersections from Open Street Maps' API and then convert it into SUMO-compatible format using netconvert. The automatically fetched area comes with an additional radius of 5 km to ensure it captures all essential connecting edges.

\subsection{Mapping Intersections and Edges}
One of the major challenges in this study was to create an algorithm to locate intersections based on the data in the SUMO network, map these intersections, extract the edges, and then map these edges to the proper direction leg. Manually finding these intersections and then mapping the edges can be time and labor-intensive and have possibilities for human error. Algorithm \ref{alg:map_intersections_edges}  is devised to streamline this process. It takes the intersection coordinates, converts them into local network map coordinates, finds the intersection, extracts the edges of this intersection, and labels each edge based on incoming and outgoing and their direction leg.

\begin{algorithm}[htbp]
\caption{Mapping Intersections and Edges}
\label{alg:map_intersections_edges}
\KwIn{net: SUMO network, target coordinates $(lon,lat)$, tolerance $tol$ (default = 5.0 m)}
\KwOut{mapping: Dictionary with incoming/outgoing edge IDs grouped by direction (east, north, west, south)}

$(x_t,y_t) \gets net.\texttt{convertLonLat2XY}(lon,lat)$\;

$min\_dist \gets \infty$\;
$closest \gets \texttt{NULL}$\;
\ForEach{node in $net.\texttt{getNodes}()$}{
    $(node\_x, node\_y) \gets node.\texttt{getCoord}()$\;
    $d \gets \sqrt{(node\_x-x_t)^2 + (node\_y-y_t)^2}$\;
    \If{$d < min\_dist$}{
        $min\_dist \gets d$\;
        $closest \gets node$\;
    }
}
\If{$min\_dist > tol$}{
    \textbf{confirm threshold (if declined, terminate)}\;
}

$junction \gets net.\texttt{getNode}(closest.\texttt{getID}())$\;

Initialize mapping: mapping[\texttt{incoming}] $\gets$ empty dictionary, mapping[\texttt{outgoing}] $\gets$ empty dictionary\;

\SetKwFunction{processEdges}{processEdges}
\SetKwProg{Fn}{Function}{:}{end}
\Fn{\processEdges{edgeList, junction, type}}{
    \ForEach{edge in edgeList}{
        \If{not (edge.\texttt{getType}() contains \texttt{highway} and not contains \texttt{footway})}{
            \tcp*[h]{Skip non-highway or footway edges.}
            \textbf{continue}\;
        }
        \uIf{$type$ equals \texttt{incoming}}{
            $(p_x,p_y) \gets edge.\texttt{getFromNode}().\texttt{getCoord}()$\;
            $angle \gets$ bearing from $(p_x,p_y)$ to $junction.\texttt{getCoord}()$\;
        }
        \Else{
            $(p_x,p_y) \gets edge.\texttt{getToNode}().\texttt{getCoord}()$\;
            $angle \gets$ bearing from $junction.\texttt{getCoord}()$ to $(p_x,p_y)$\;
        }
        $dir \gets \texttt{classify}(angle)$\;
        Append edge.\texttt{getID}() to mapping[$type$][$dir$]\;
    }
}

\processEdges{$junction.\texttt{getIncoming}()$, junction, \texttt{incoming}} \;
\processEdges{$junction.\texttt{getOutgoing}()$, junction, \texttt{outgoing}} \;

\Return mapping\;

\tcc{%
  \textbf{classify(angle)}: Returns the cardinal direction: \\
  \quad \texttt{east}: if $angle \geq 315^\circ$ or $angle < 45^\circ$ \\
  \quad \texttt{north}: if $45^\circ \leq angle < 135^\circ$ \\
  \quad \texttt{west}: if $135^\circ \leq angle < 225^\circ$ \\
  \quad \texttt{south}: if $225^\circ \leq angle < 315^\circ$
}
\end{algorithm}

\subsection{Converting Turn Counts to Vehicle Flows}
\label{generation_methods}
With the SUMO network in place, the next step is to use the turn count data to create vehicle demand in SUMO. Demand generation can be done using two preferred methodologies for this case. The first one uses a route-creation method \cite{shokrolah2022turning} where explicit vehicle routes are created based on a start and end point on the map, where the intersection of interest lies somewhere between these routes. The other method uses flow definition to generate traffic flow, and the start and end points for vehicles following a flow generation will be incoming and outgoing edges, respectively. Table \ref{table_flow_vs_vehicle} compares both methodologies.
Since the available TMC data is aggregated into 15-minute bins, the flow generation technique is optimal for our case while keeping the complexity of our scenario to a minimum and making it simple to calibrate and validate.

CrossFlow is designed to automatically assign approach flows in the proper direction, along with the right vehicle type and vehicle movement. For example, if the northbound-left movement has 150 cars per hour, we create a flow of 150 cars entering from the north incoming edge and turning left at the junction. The departure times of vehicles in that flow are uniformly distributed over 1 hour to avoid dispatching all of them simultaneously. We divide the 1-hour flow into smaller intervals that match our data, 4 divisions in this case, and then assign flows for finer granularity. This approach also aligns with SUMO’s support for using count data on junctions to generate flows.

\begin{table*}[ht]
    \caption{Comparison of Flow Definitions and Explicit Vehicle Routes in SUMO}
    \label{table_flow_vs_vehicle}
    \centering
    \begin{tabular}{|>{\centering\arraybackslash}m{4cm}|>{\centering\arraybackslash}m{4cm}|>{\centering\arraybackslash}m{4cm}|}
        \hline
        \textbf{Feature} & \textbf{Flow Definitions} & \textbf{Explicit Vehicle Routes} \\
        \hline
        Representation & \texttt{<flow>} tag & \texttt{<vehicle>} tag \\
        \hline
        Granularity & Aggregated over time & Individual vehicle level \\
        \hline
        File Size & Smaller & Larger \\
        \hline
        Control Over Departures & Uniform or randomized timing & Precise timing per vehicle \\
        \hline
        Ideal For & Large-scale network simulations & Detailed per-vehicle analysis \\
        \hline
        Ease of Use & Simple to generate and edit & More complex but flexible \\
        \hline
    \end{tabular}
\end{table*}

\subsection{Simulation Parameters}
\label{simparameters}
SUMO allows the configuration of several parameters in the simulation. While our tool defaults these parameters to match standard values, it still allows users to modify them as per their needs.
\begin{itemize}
    \item Car Following Model - is responsible for how vehicles interact with other vehicles by adjusting parameters like speed and maintaining a certain distance from other vehicles. By default, SUMO uses a modified version of the Krauss Car Following model. CrossFlow allows the user to select other models as well, including but not limited to the original Krauss Model \cite{krauss1998microscopic}, IDM\cite{Treiber2000}, and Wiedemann\cite{wiedemann1974simulation}.
    \item Vehicle Length- each vehicle type has a different length; for the current simulated scenarios, we focused on simulating cars, trucks, and buses and assigned each vehicle type a default length, which can be adjusted by the user in the tool.
    \item Driver Behavior  - denoted by the variable sigma in SUMO, it controls and defines the amount of randomness or imperfection in a vehicle's behavior. It ranges from 0 to 1, and 0 implies that the drivers behave perfectly—there is no randomness. This is deterministic behavior. Anything greater than 0 means the drivers have imperfections and act more human-like, i.e., less predictable.
\end{itemize}

\subsection{Simulation Execution}

\begin{figure}[b]
    \centering
    \includegraphics[width=\columnwidth]{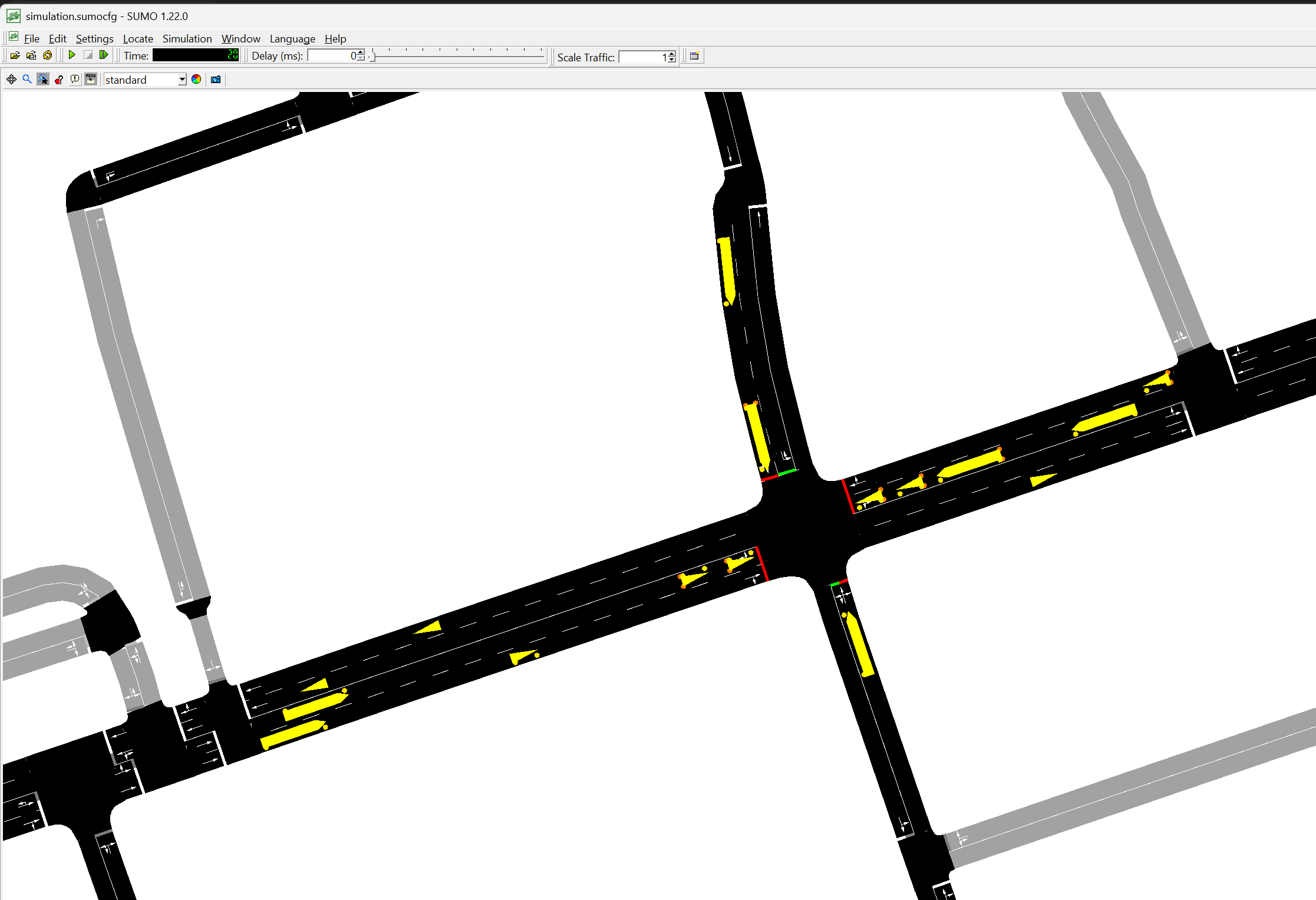}
    \caption{Intersection Simulation in SUMO.}
    \label{fig:sumo_run}
\end{figure}
Once the network and the flow files are prepared, CrossFlow generates a SUMO configuration file, which links the network and the vehicle flow definition file to run the simulation. The tool can run the simulation directly if the user prefers not to open the configuration file in SUMO manually. For quick run and visual verification, the tool can launch SUMO-GUI with the generated scenario so the user can see vehicles moving through the intersection in real-time. This helps confirm that vehicles follow expected paths and that the volumes qualitatively look correct. Figure \ref{fig:sumo_run} shows vehicle demand generation at an intersection created using CrossFlow.

\section{Tool Architecture and Features}\label{Tool}
      \begin{figure}[tb]
    \centering
    \includegraphics[width=0.95\columnwidth]{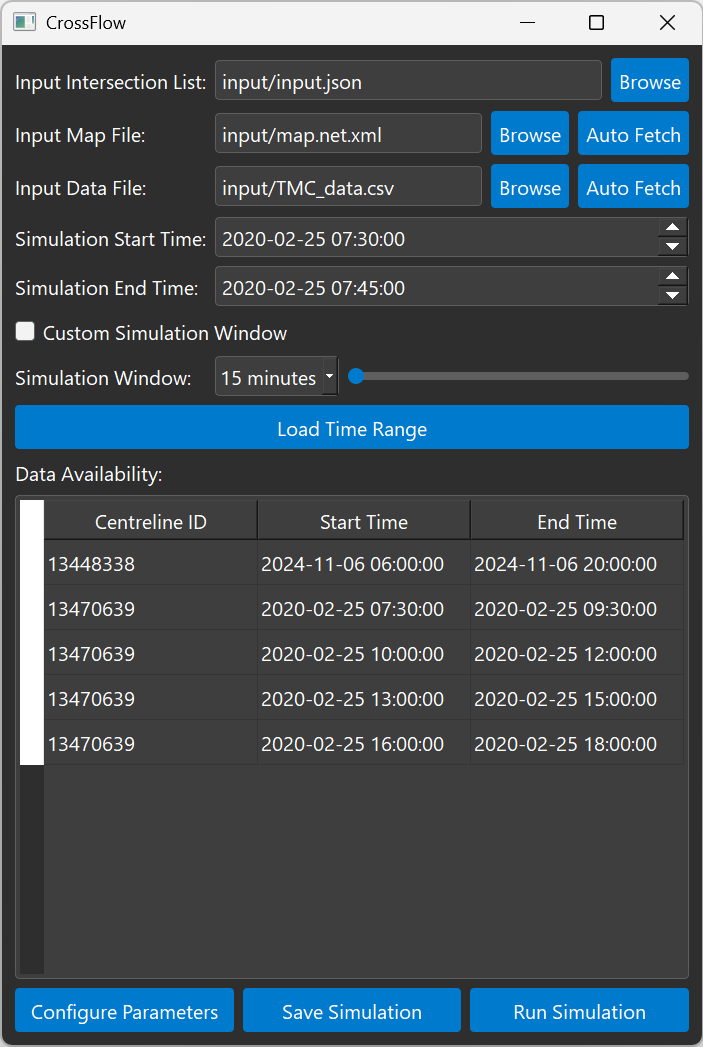}
    \caption{GUI Demonstration of CrossFlow.}
    \label{fig:gui_main}
\end{figure}

\begin{figure}[H]
    \centering
    \includegraphics[height=12cm]{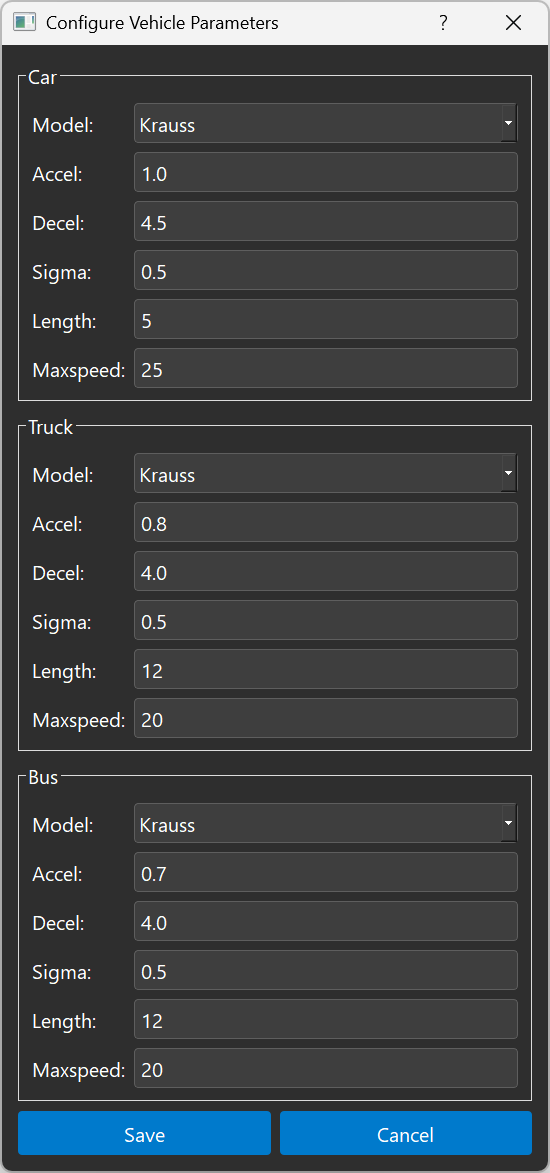}
    \caption{Edit Vehicle Configurations using The Tool's GUI.}
    \label{fig:gui_config}
\end{figure}

The developed tool is implemented in Python and encapsulates the above methodology into a user-friendly application. Figure \ref{fig:gui_main} shows a screenshot of the tool’s GUI. The tool's architecture can be conceptually divided into three components: the GUI front-end, the simulation engine integration, and the data processing backend. The tool makes running a traffic simulation easy for anyone with basic computing knowledge; the process of running a simulation using CrossFlow is a five-step process:

\begin{enumerate}[label=Step \arabic*:, leftmargin=*, labelsep=1em]
    \item \textbf{Input a list of Centerline IDs} of the intersections the user wants to simulate. The list should be in a simple comma-separated JSON format. Centerline IDs are unique location identifiers corresponding to the Toronto Centreline or Intersection File datasets.
    
    \item \textbf{Input the SUMO-compatible Map}, which is detailed in Section~\ref{buildMap}. Additionally, we have incorporated OpenStreetMap's API to download the map spanning the intersections and convert it to SUMO-compatible format, enabling the user to have the map ready with just a single click.
    
    \item \textbf{Load the TMC data file}, for which the user can browse manually if they have the data available locally or use the \textit{Auto Fetch} feature. This feature queries the City of Toronto's Open Data Portal API to load the data into the tool automatically.

    \item \textbf{Select Start and End Time}. The \textit{Load Time Range} feature processes all the data against the selected Centerline IDs to identify available data that can be used to simulate traffic. Users can choose from predefined simulation times (15, 30, 45, or 60 minutes) or specify a custom simulation window for an extended simulation.

    \item \textbf{[Optional] Configure Parameters}. This feature allows users to change various parameters used during the simulation, as shown in Figure~\ref{fig:gui_config} and discussed in detail in Section~\ref{simparameters}.
    
    \item \textbf{Run the Simulation}. Once all inputs are configured, the user simply clicks \textit{Start Simulation}, and the system saves the SUMO configuration and the vehicle routes file and opens SUMO-GUI, which looks like Figure \ref{fig:sumo_run}. Alternatively, the user can save the simulation at the moment and run it later using the configuration file in SUMO or using other methods.
\end{enumerate}

This tool was created with extensibility in mind. The source code is made public for this tool, making it feasible for anyone to implement new or improve existing functions. While the current focus is on Toronto intersections, the underlying approach can work for any TMC data:
\begin{itemize}
    \item The data import module is made adaptable and will automatically adapt to different time interval bins, as some cities also record data in the 5-minute, hourly, or custom range.
    \item The network generation is not limited to just a four-way typical intersection; it is tested with 3-way intersections as well. While other intersections, like roundabouts or specialized ones, were not included in the scope of this study, support can be extended with a few modifications.
    \item The tool can be extended to incorporate bikes and pedestrian data to simulate non-vehicular movements through the intersection. This can be useful for studying pedestrian timing or safety, but our main evaluation is on vehicular volumes for simplicity.
\end{itemize}
Finally, the GUI includes a "Save Simulation" feature, enabling users to export the SUMO configuration files. This transparency helps users monitor the vehicle flow and variables and further tweak the scenario outside the tool if needed. For instance, custom traffic light timings can be configured externally via the SUMO configuration file to test, or the configuration file can be used alongside TraCI (SUMO's programming API) to analyze and modify the network's parameters in real-time.

This tool offers a one-click import feature to gather the latest data and prepare the intersection map without manual intervention while remaining highly adaptable and extensible. The flexible data module accommodates various time interval bins, and the network generation can handle various intersection types naively, with a potential for more complex layouts with slight modifications. The user-friendly GUI further simplifies the process with a one-click “Save Simulation” feature, enabling seamless export without requiring advanced programming skills.

\section{Evaluation}\label{Evaluation}

\begin{figure}[tb]
    \centering
    \includegraphics[width=\columnwidth]{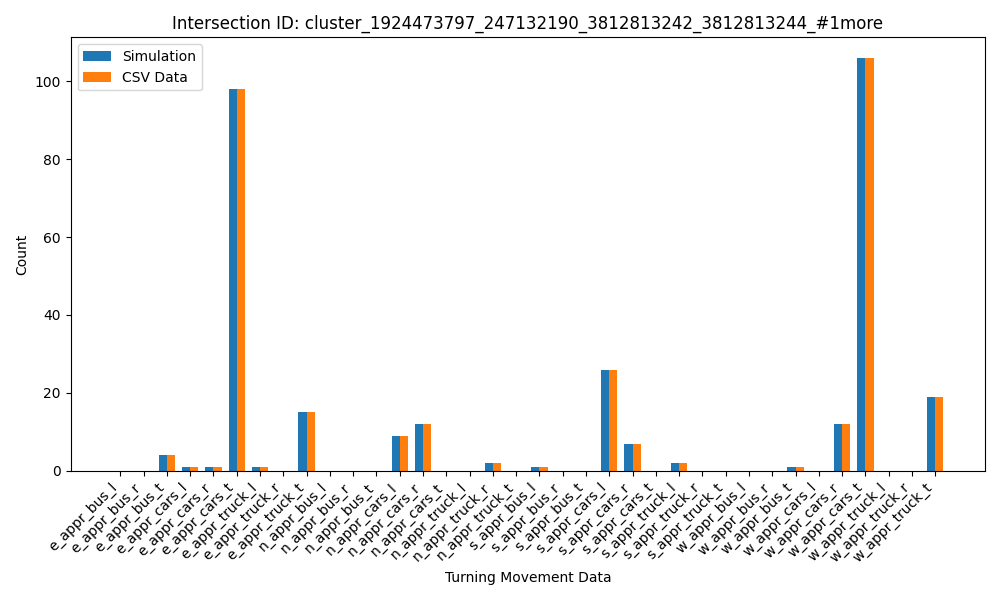}
    \caption{Simulated vs Real Traffic Count for One of the Intersections.}
    \label{fig:chart1}
\end{figure}

The evaluation compares simulated vehicle movement in SUMO with real TMC data from the City of Toronto. The SUMO configuration file created using our tool is run using TraCI (Traffic Control Interface), which comes prepackaged with the simulator. TraCI allows retrieving individual component states and values in SUMO and supports modifying them online. If the simulation runs for 15 minutes, it will span 900 simulation steps. Vehicle IDs are collected from the monitored incoming and outgoing edges of the intersection, where traffic is simulated for each simulation step. Each vehicle ID is parsed to extract a turning movement key, and unique vehicle occurrences are recorded per intersection. A new TMC is created using the simulated Intersection traffic volume, which is then compared against the real data. This systematic process validates the simulation by highlighting any discrepancies between the modeled and real-world traffic counts. We tested our results for 25 different intersections, some individual, while some were simultaneously simulated, and concluded that our simulation technique effectively simulates any regular 3- or 4-way intersection, and the simulated vehicle count matched with real data. Figure \ref{fig:chart1} shows one such evaluation of an intersection in Toronto, which is simulated for 15 minutes. The high effectiveness of this traffic generation method arises from the inherent tradeoff of limiting control over vehicles at an individual scale, as discussed in Section \ref{generation_methods}.

\section{Use Cases and Applications}\label{Use}

The ability to virtually replicate real-world traffic scenarios unlocks various practical applications in traffic engineering and research. In this section, we outline a few use cases of how the tool can be employed to address common intersection analysis questions:

\subsection{Traffic Signal Timing Optimization}
One immediate application for the scenarios generated using this tool is to apply traffic signal optimization techniques, as discussed in Section \ref{Introduction}; the modern TSC techniques leverage realistic virtual environments to improve their accuracy in the real world. Additionally, the effect of adding or modifying phases can also be tested. For instance, an intersection with no dedicated phase left turn can be modified to have a protected left-turn phase and see the impact if it clears left-turn queues more effectively with a minimal trade-off with other vehicle lane queues.

\subsection{Scenario Analysis with Volume Changes}
Future planning or what-if scenarios are one of the major tasks in transportation engineering. Because our tool allows easy manipulation of the volume inputs not directly, one can multiply the data by a certain percentage, which is available in a CSV format, and simulate scenarios to check the operating capacity of an intersection with more or less traffic volume. SUMO natively supports scaling the traffic during simulation, which is also helpful. Similarly, transportation mode volumes can be manipulated to check the impact on urban traffic with fewer cars and more public transport vehicles like buses.

\subsection{Intersection Design and Control Changes}
The tool can assist city planners in assessing physical design changes or control changes. For instance, adding or removing a turn lane depends on the turn volumes at an intersection. Adding a dedicated turn lane if the turn volume is high or replacing a dedicated turn lane if the turn volume is minimal can significantly impact traffic flow. Another scenario is where lower traffic volume intersections with traffic signals can be tested by replacing signals with stop signs and monitoring the impact on traffic flow. Various scenarios can be tested using the same vehicle flows generated using CrossFlow and modifying the network file.

\subsection{Academic and Research Extensions}
Another use case for this tool is in the educational demonstration; the intuitive GUI and visual output make it a valuable tool for students learning traffic engineering concepts; they can quickly input real data, modify data, create custom data, run scenarios, and learn how traffic in an intersection behaves. Reinforcing concepts like how volume and green time interplay can also be studied. Furthermore, this tool can be extended to various research areas, like calibrating simulation models to create a feedback loop, further reducing the gap in real and simulated vehicular interactions. Evaluation of new techniques for urban traffic can be done more robustly as realistic scenarios create better testbed environments with human-like variables that randomly generated data cannot come close to.

\section{Conclusion}\label{Conclusion}
This paper presented a detailed look at urban traffic simulations focused on intersections that use real-world turn counts as inputs to the SUMO microsimulator. CrossFlow streamlines the process of converting raw empirical data into SUMO-compatible vehicle flows. Key findings from our research point to the feasibility of using real-world data in urban simulations. In the broader context of transportation research, our work contributes to the ongoing traffic modeling efforts incorporating big data and open data. We apply one of many possible approaches to put the data to use in microsimulation modeling for improved analysis and planning. It corresponds with the findings discussed in the earlier section about realistic simulation requiring realistic inputs and providing a concrete implementation targeting intersections, which frequently serve as the initial level of analysis in urban traffic studies. Future work includes enhancing the tool by integrating one-click import functionality for additional cities that provide TMC data globally. Furthermore, pedestrian and bicycle data will be combined along with the vehicular volume to extend the applicability of the system. Lastly, the development of a route-based traffic flow generation method is also planned, aiming to provide more detailed and accurate modeling capabilities and individual vehicle control in the simulation.
In conclusion, we have demonstrated that data-driven intersection simulation is practical and beneficial. Using TMC data, CrossFlow provides a credible simulation environment that can help analyze traffic patterns, find weak links in the intersection network, or use the environment as a test bed to develop and enhance traffic optimization techniques in a low-cost, risk-free, and efficient way. As cities continue to collect and release traffic data, tools like the one presented can play a vital role in turning data into actionable insights for more intelligent and safer transportation systems.

\section*{Acknowledgment}

This work is partially sponsored by the Natural Sciences and Engineering Research Council of Canada (NSERC) through the Discovery Grants (DG) program.

\input{conference_101719.bbl}

\vspace{12pt}

\end{document}